\documentclass[a4paper,fleqn]{cas-dc}

\usepackage[numbers,sort&compress]{natbib}
\usepackage{siunitx}
\usepackage{microtype}

\def\tsc#1{\csdef{#1}{\textsc{\lowercase{#1}}\xspace}}
\tsc{WGM}
\tsc{QE}
\tsc{EP}
\tsc{PMS}
\tsc{BEC}
\tsc{DE}

\begin{document}
\let\WriteBookmarks\relax
\def\floatpagepagefraction{1}
\def\textpagefraction{.001}

\shorttitle{EIGNN}

\shortauthors{Xiao jiang et~al.}

\title [mode = title]{Enhancing the Prediction of Glass Dynamics by Incorporating the Direction of Deviation from Equilibrium Positions}                      



%
\author[1]{Xiao Jiang}[]


\affiliation[1]{organization={College of Computer Science and Electronic Engineering,},
    addressline={Hunan University}, 
    city={Changsha},
    postcode={410012}, 
    country={China}}

\author[1]{Zean Tian}[]
\cormark[1]


\ead{tianzean@hnu.edu.cn}

\author[1]{Kenli Li}[%
   ]
\author[2]{Wangyu Hu}[%
   ]


\affiliation[2]{organization={College of Materials Science and Engineering},
    city={Changsha},
    postcode={410012}, 
    country={China}}

\cortext[cor1]{Corresponding author. }

\begin{abstract}
Elucidating the intricate relationship between the structure and dynamics in the context of the glass transition has been a persistent challenge.  Machine learning (ML) has emerged as a pivotal tool, offering novel pathways to predict dynamic behaviors from structural descriptors. Notably, recent research has highlighted that the distance between the initial particle positions between the equilibrium positions substantially enhances the prediction of glassy dynamics. However, these methodologies have been limited in their ability to capture the directional aspects of these deviations from the equilibrium positions, which are crucial for a comprehensive understanding of the complex particle interactions within the cage dynamics. Therefore, this paper introduces a novel structural parameter: the vectorial displacement of particles from their initial configuration to their equilibrium positions. Recognizing the inadequacy of current ML models in effectively handling such vectorial parameters, we have developed an Equivariance-Constrained Invariant Graph Neural Network (EIGNN). This innovative model not only bolsters the descriptive capacity of conventional rotation-invariant models but also streamlines the computational demands associated with rotation-equivariant graph neural networks. Our rigorous experimental validation on 3D glassy system from GlassBench dataset has yielded compelling evidence that the EIGNN model significantly enhance the correlation between structural representation and dynamic properties. 
\end{abstract}



\maketitle

\section{Introduction}
In recent years, with the rapid development of big data analytics, machine learning, and neural networks, establishing effective machine learning models to address physical problems in the complex glassy system is currently a research hotspot\cite{clegg2021characterising,liu2022challenges,berthier2023modern,cubuk2015identifying,bapst2020unveiling, wang2021inverse,wang2024predicting}. For example, the development of machine-learned "softness" \cite{cubuk2015identifying,cubuk2017structure,schoenholz2017relationship} has resolved a long-standing puzzle in physics: the existence of structural signatures correlated with dynamics during the glass transition. This discovery has catalyzed a wave of research, broadening our understanding of glassy phenomena across various liquids and disordered solids \cite{sussman2017disconnecting,harrington2019machine,liu2019yiruan,ma2019heterogeneous,cubuk2020unifying,wuyicheng2021,ridout2022correlation,tah2022fragility,liu2022deciphering}.

In 2020, the DeepMind team \cite{bapst2020unveiling} revolutionized glassy research by introducing GNN-DM, a graph neural network (GNN) that surpasses traditional machine learning approaches. After that, studies have refined the GNN-DM model\cite{shiba2023botan,pezzicoli2024rotation, jung2023predicting,alkemade2023improving,jiang2023geometry}. The bond-targeting network (BOTAN) \cite{shiba2023botan} employs a specialized loss function that evaluates single-particle mobility and pairwise motion. The SE(3)-equivariant model\cite{pezzicoli2024rotation} incorporates geometric rotation-equivariance and refines node features with potential energy for improved long-time forecasts. GlassMLP \cite{jung2023predicting} marks the latest advancement, utilizing Voronoi volumes and descriptors such as particle potential energy to forecast dynamic heterogeneity at the glass transition temperature $T_g$\cite{jung2024dynamic}. This model also introduces the innovative application of inherent structure.



A significant advancement in this field is the incorporation of the cage state structure, which represents the equilibrium positions around which particles vibrate when no particle rearrangement occurs \cite{alkemade2023improving}. By utilizing Monte Carlo simulations to determine the cage center positions, this method innovatively introduces the distance parameter between the initial and equilibrium configurations. This parameter serves as input for machine learning models, resulting in improved predictions of particle dynamic propensity during the initial stages of cage times and relaxation times, surpassing existing graph-based deep learning approaches.

However, current methods for characterizing deviations from equilibrium positions focus solely on distance and neglect directional aspects\cite{jung2023roadmap}. To address this limitation, this paper introduces the displacement vector from the initial particle structure to the equilibrium position as a crucial structural parameter. To effectively encode this parameter and the geometric structure, we have developed the Equivariance-Constrained Invariant Graph Neural Network (EIGNN) to enhance predictions of glassy dynamics.


\section{Dataset and Displacement parameter }

\subsection{Dataset}
To validate the effectiveness of the proposed model, this paper utilizes the relaxation dynamics data of 3D glass-forming liquids from the GlassBench dataset\cite{jung2023roadmap}, rwhich is referred to as GlassBench-3D for simplicity. This dataset, introduced by Shiba et al.\cite{shiba2023botan}, includes relaxation dynamics data at four distinct temperatures: T=0.44, T=0.50, T=0.56, and T=0.64. The simulation data focuses on the 80:20 Kob-Andersen (KA) mixture, consisting of two particle types, A and B, interacting via the Lennard-Jones potential\cite{kob1995testing}. 

For each temperature, the dataset provides 500 configurations ${\{\mathbf{\vec{r}}_{i}^{\rm init}\}}_{i = 1...N_p}$ along with their corresponding dynamic labels. Moreover, the GlassBench-3D data includes the inherent structure ${\{\mathbf{\vec{r}}_{i}^{\rm inh}\}}_{i = 1...N_p}$ for all initial configurations and cage structure ${\{\mathbf{\vec{r}}_{i}^{\rm cage}\}}_{i = 1...N_p}$ for 200 configurations.
For each initial configuration in the dataset, 30 micro-canonical simulations are run independently,  with initial velocities sampled from the Maxwell-Boltzmann distribution. During these simulations, particle configurations are recorded at logarithmically spaced intervals from $t=0.13$ to $t=130000$. 

The dynamic labels epresent the propensity of each particle\cite{berthier2007structure, widmer2004reproducible, propensity1}. The propensity $y_i$ of particle $i$ over a time $ t $ is defined as the average absolute displacement over the $N_{\rm run} =32$ simulations runs:
\begin{equation}
y_i(t) = \langle \lVert \mathbf{\vec{r}}_i(t) - \mathbf{\vec{r}}_i^{\rm init}\rVert \rangle_{\rm iso},
\end{equation}
where 'iso' denotes the iso-configurational ensemble average, $\mathbf{\vec{r}}_i(t)$ is the position of particle $i$ at time t.

\subsection{Displacement parameter}

With the cage effect, the particles are constrained to oscillations around their equilibrium positions in the short times.
The unique cage dynamics mechanism suggests that the distance $ d_i^{\rm equi} = \lVert\mathbf{\vec{r}}_i^{\rm equi} -  \mathbf{\vec{r}}_i^{\rm init}\rVert$ between a particle's initial position $ \mathbf{\vec{r}}_i ^{\rm init}$ and the equilibrium position $\mathbf{r}_i^{\rm equi} $, as a structural parameter, can effectively reflect short-time propensity and further enhance long-time propensity prediction\cite{alkemade2023improving}. Now, a pertinent question is whether the displacement $ \mathbf{\vec{d}}_i^{\rm equi}  = \lVert\mathbf{\vec{r}}_i^{\rm equi} -  \mathbf{\vec{r}}_i^{\rm init}\rVert$ between the initial position and the equilibrium position can improve the prediction of dynamics. 


Next, we use the framework of equivariant graph neural networks to validate the effectiveness of these parameters for dynamic prediction.

\begin{figure*}[htp]
    \centering
    \includegraphics{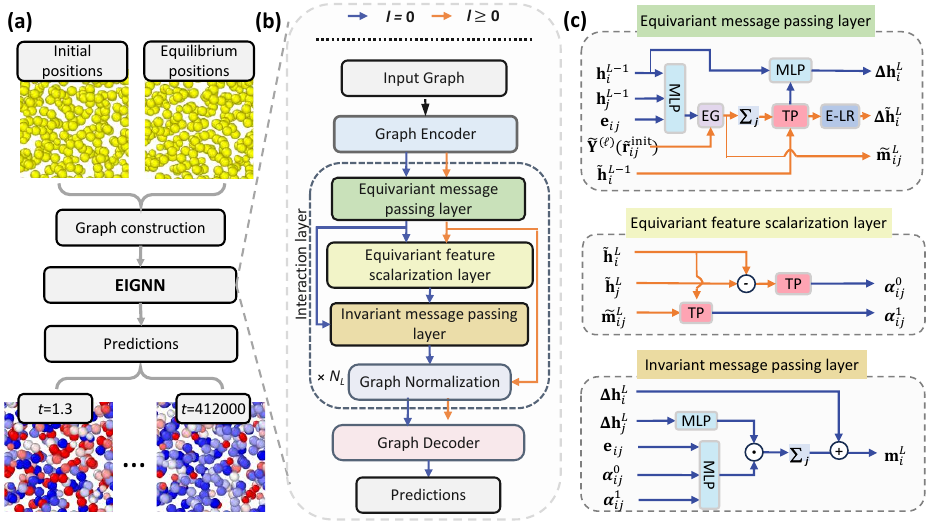}
    \caption{\textbf{(a)} The workflow of our proposed propensity prediction framework based on EIGNN. \textbf{(b)} The architecture of the EIGNN model. \textbf{(c)} The network of the sub-layers in the interaction layer. TP denotes the tensor product, EG denotes the equivarint gate, and E-LR represents the equivariant linear layer.  }
    \label{fig:model}
\end{figure*}
\section{Methods}
Figure \ref{fig:model}(a) presents the workflow of our propensity prediction framework.
This paper proposes a comprehensive graph representation that integrates the initial structure, equilibrium structure, and displacement parameters in constructing the graph structure. 

After building the graph, existing equivariant graph neural network (GNN) models can be used to characterize these equivariant features and predicting the dynamics. However, strictly equivariant GNNs \cite{pezzicoli2024rotation, thomas2018tensor,batzner20223,brandstetter2022geometric,batatia2022mace,liao2023equiformer} have high computational complexity, especially as the rotational order $\ell_{max}$ of the equivariant features increases, causing a sharp rise in the computational cost of tensor products. Thus, designing efficient equivariant convolution operations to reduce computational costs is a primary goal for targeted equivariant GNNs \cite{passaro2023reducing,luo2024enabling}. However, their computational costs remain higher than GNNs based on rotationally invariant features, limiting their further application in large-scale glassy system. Therefore, this paper proposes an efficient Equivariance-Constrained Invariant Graph Neural Network (EIGNN) to learn the glassy structural representation. 


\begin{figure}
    \centering
    \includegraphics{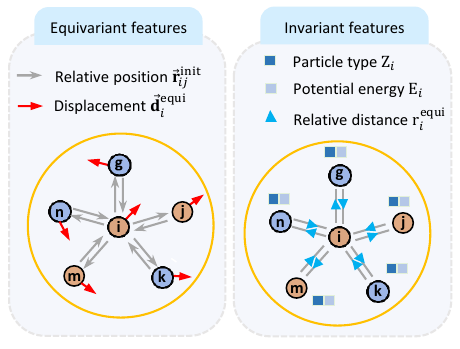}
    \caption{Illustration of the equivariant  and invariant features used in the graph. }
    \label{fig:graph}
\end{figure}
\subsection{Graph construction}

First, the particle neighbor relationships $\mathcal{E}^{init}$ and $\mathcal{E}^{\rm equi}$ corresponding the two input structures are separately determined based on a cutoff distance $r_c =2$.  Next, the edges $(i,j$) in the graph are determined by the union $\mathcal{E}^{init} \cup \mathcal{E}^{\rm equi}$. We observe a substantial overlap of neighboring particles between the two structures, indicating that many particles are adjacent in both configurations. The number of edges in the merged input graph does not significantly increase. 

The node and edge features of the input graph are shown in Fig.\ref{fig:graph}. Here, the rotation-equivarint features and the invariant features are separately shown. The equivaint features of graph include the relative position $\mathbf{\vec{r}}_{ij}^{rm init}$ of particle pair $ij$ in the initial structure and the displacement vector  $\mathbf{\vec{d}}_i^{\rm equi}$ of node $i$. The invariant features of graph include several aspects: (1) the particle type $Z_{i}\in{0,1}$ to distinguish particle types in the KA system; (2) the potential energy $E_i = \sum_{j\neq i} V(r_{ij})/2$, reflecting the potential energy of the particle; the relative distance $\lVert\mathbf{\mathbf{\vec{r}}}^{equi}_{i} - \mathbf{\mathbf{\vec{r}}}^{equi}_{j} \rVert$ in the equilibrium structure. 

With this construction strategy, the generated geometric graph can comprehensively cover the core information of both structures.
Note that the orientation of edge $ij$ is omitted, as it can be infer from the $\mathbf{\vec{d}}_i^{\rm equi}$ and $\mathbf{\vec{d}}_j^{\rm equi}$.

\subsection{Architecture of EIGNN}

The architecture of the proposed EIGNN model is illustrated in Fig. \ref{fig:model}. It consists of a graph encoder, interaction layers, and a final decoder layer. Each interaction layer comprises an equivariant message-passing layer, an equivariant interaction layer, an invariant message-passing layer, and the graph normalization. 

The EIGNN interaction layer employs two distinct latent feature spaces : one is the rotationally invariant feature space containing scalar features ($\ell=0$), and the other is the rotationally equivariant feature space that handles geometric tensors of any rotational order $\ell \geq 0$. By constraining the message passing of rotationally invariant features through interactions with equivariant features, the computational complexity of the EIGNN is significantly reduced compared to traditional equivariant message-passing networks.

The following sections provide a detailed description of each layer in the EIGNN.

\textit{\textbf{Graph encoder}}
In handling the SO(3) equivariance in geometric graphs, appropriate initialization of parameters on nodes and edges is crucial. The displacement vector $\mathbf{\tilde{d}}_{i}^{\rm equi}$ on the node are decomposed into the norm $d_i^{\rm equi} = \lVert\mathbf{\vec{d}}_{i}^{\rm equi} \rVert$ and orientation $\mathbf{\tilde{d}}_{i}^{\rm equi} = {\mathbf{\vec{d}}_{i}^{\rm equi}}/{d_i^{\rm equi}}$. Similarly,  $\mathbf{\tilde{r}}_{ij}^{\rm init}$ are decomposed into $r_{ij}^{\rm init}=\lVert\mathbf{\vec{r}}_{ij}^{\rm init}\rVert $ and $\mathbf{\tilde{r}}_{ij}^{\rm init} = \mathbf{\tilde{r}}_{ij}^{\rm init} / r_{ij}^{\rm init}$.

The invariant features of node $i$ are initialized through the following linear layer:
\begin{equation}
\mathbf{h}_i^{L=0} = \mathbf{W}_i^{L=0}\cdot\left[f_{\text{1HoT}}(Z_i)\oplus{E_i}\oplus{}d_i^{equi}\right]
\end{equation}
where $\mathbf{W}_i^{L=0}$ is the weight matrix of the linear layer, $f_{\text{1HoT}}$ denotes the one-hot encoding of the particle type, and $\oplus$ indicates vector concatenation.

The initial equivariant features of node $i$ are defined as a combination of geometric tensors \cite{thomas2018tensor} corresponding to different rotational orders $\ell\in{0,...,\ell_{\text{max}}}$. Mathematically, the node features can be expressed as:
\begin{equation}
{\mathbf{\tilde{h}}_i^{L=0}} = [\underbrace{{{h}}_i^{(0),L=0}}_{\mathbb{R}^{1}},..., \underbrace{{\mathbf{\tilde{h}}}_i^{(\ell_{max}),L=0}}_{\mathbb{R}^{2\ell_{max}+1}}]\in\mathbb{R}^{{(\ell_{max} + 1)^2},}
\end{equation}
where the geometric tensor ${\mathbf{\tilde{h}}i^{(\ell),L=0}}$ for each rotational order $\ell$ is initialized using the $\ell$-th order spherical harmonics $\tilde{Y}^{(\ell)}(\mathbf{\hat{d}}{i})$ of the parameter $\mathbf{\hat{d}}_{i}$. Unlike existing equivariant graph neural networks \cite{brandstetter2022geometric, pezzicoli2024rotation, geiger2022e3nn} that set multiple channels for different rotational orders, this work balances computational efficiency and performance by separating the equivariant and invariant feature spaces. A single-channel feature is used in the equivariant feature space ($\ell \geq 0$), while the dimensionality of features is increased in the scalar feature space ($\ell = 0$).

To ensure the stability of the training process, the rotationally invariant parameters $r_{ij}^{\rm equi}$ and $r_{ij}$ on edge $(i,j)$ are expanded into $N_b$-dimensional vectors $\mathbf{B}(r_{ij}^{\rm equi})$ and $\mathbf{B}(r_{ij}^{\rm init})$ via radial basis functions $B_{n\in{1,...,N_b}}(r)$ before being passed to the structural representation layer:
\begin{align}
&\mathbf{B}(r_{ij}^{\rm equi}) =[B_0(r_{ij}^{\rm equi}) \oplus...\oplus B_{N_{b}}(r_{ij}^{\rm equi}) ] \\
&\mathbf{B}(r_{ij}^{\rm init}) =[B_0(r_{ij}^{\rm init})\oplus...\oplus B_{N_{b}}(r_{ij}^{\rm init}) ].
\end{align}
The adopted radial basis function is the Bessel basis function proposed by \cite{klicpera2020directional}, which has a polynomial envelope function:
\begin{equation}
B_n(r) = \sqrt{\frac{2}{r_c}}\frac{\sin\left(\frac{n\pi}{r_c}r\right)}{r},
\end{equation}
where $r_c$ is the cutoff distance. 

Finally, the initial feature representation of edge $(i,j)$ is $\mathbf{e}{ij} = [\mathbf{B}(r_{ij}^{\rm equi}) \oplus {\mathbf{B}(r_{ij}^{\rm init})}]$. Additionally, orientation $\mathbf{\tilde{r}}_{ij}^{\rm ini}$ is mapped as a combination of spherical harmonics $\tilde{\mathbf{Y}}^{(\ell)}(\mathbf{\tilde{r}}_{ij}^{\rm init})$ for different rotational orders $\ell\in{0,...,\ell_{\text{max}}}$. These initializations, as fundamental elements of the equivariant model, ensure that the model can effectively handle the rotationally equivariant parameters in geometric graphs.

\textit{\textbf{Equivariant message passing layer}}
First, this layer uses a linear combination of spherical harmonics to lift scalar features ($\ell=0$) to geometric tensors of any rotational order $\ell \geq 0$. The weights of the linear combination are derived from the atomic scalar features and radial density function estimation. The specific steps are as follows:
\begin{align}
&m_{ij}^{\ell, L} = \text{MLP}\left(\left[ \mathbf{h}i^{L-1} \oplus {\mathbf{e}_{ij}} \oplus {\mathbf{h}j^{L-1} }\right]\right) \label{short1} \\
&\mathbf{\tilde{m}}_{ij}^{(\ell),L}= {m_{ij}^{\ell, L}\cdot{\tilde{\mathbf{Y}}^{(\ell)}(\mathbf{\tilde{r}}_{ij}^{\rm init})} } \label{short2} \\
&{\tilde{\mathbf{m}}}i^{(\ell),L} = {\sum_{j\in{\mathcal{N}_i}}\mathbf{\tilde{m}}_{ij}^{(\ell),L}} .\label{short3}
\end{align}
Equations \ref{short1} and \ref{short2} introduce nonlinearity to the spherical harmonics of rotational order $\ell$ using an equivariant gating activation mechanism. The MLP first acts on the scalar features to obtain nonlinear weights, using a sigmoid activation function to convert the scalar features into nonlinear weights between 0 and 1. These weight coefficients $m{ij}^{\ell, L}$ are then multiplied by the corresponding spherical harmonics to obtain the equivariant features $\mathbf{\tilde{m}}_{ij}^{(\ell),L}$ on the edges. Finally, Equation \ref{short2} aggregates the equivariant features from the neighborhood to obtain the equivariant message ${\tilde{\mathbf{m}}}_i$ for node $i$. The interaction process is completed by the Clebsch-Gordan (C-G) tensor product.

Subsequently, the equivariant messages of the nodes are further fused with the node features from the previous layer. The specific steps are as follows:
\begin{align}
   & \Delta\mathbf{\tilde{h}}_i^{({\ell_1,\ell_2})\to(\ell_3),L} = {\tilde{\mathbf{m}}}_i^{(\ell_1),L} \mathop{\otimes}\limits_{{\ell_1,\ell_2}}^{\ell_3}{\tilde{\mathbf{h}}}_i^{(\ell_2),L-1} \label{emb1}\\
   & \Delta\mathbf{m}_i^{L} =\left[\mathop{\oplus}\limits_{({\ell_1,\ell_2})} \Delta\mathbf{\tilde{h}}_i^{({\ell_1,\ell_2})\to(\ell_3 = 0),L} \right]\\
    &\Delta\mathbf{h}_i^{L} = \text{MLP}\left(\left[\Delta\mathbf{m}_i^{L} \oplus{\mathbf{h}_i^{L-1} }\right] \right) \label{emb2}\\
   & \mathbf{\tilde{h}}_i^{(\ell),L}  = \sum_{({\ell_1,\ell_2})}{w^{({\ell_1,\ell_2})\to(\ell), L} }\cdot \Delta\mathbf{\tilde{h}}_i^{({\ell_1,\ell_2})\to(\ell),L} \label{emb3}
\end{align}

Equation \ref{emb1} calculates the C-G tensor product between the equivariant features from the previous layer and the current equivariant messages of the nodes. The equivariant message ${\tilde{\mathbf{m}}}_i^{(\ell_1),L}$ of rotational order $\ell_1$ and the equivariant features ${\tilde{\mathbf{h}}}_i^{(\ell_2),L-1}$ of rotational order $\ell_2$ generate features of rotational order $\ell_3$ according to the path constraint $|\ell_1 - \ell_2| \le \ell_3 \le |\ell_1 +\ell_2|$. Additionally, an extra constraint $\ell_3<\ell_{max}$ is introduced in the calculation to limit the order of the output equivariant features, thereby preventing an increase in dimensions and improving computational efficiency.

In Equation \ref{emb2}, the scalar output of the tensor product is further combined with the scalar features of the previous layer's nodes. The operator $\oplus_{({\ell_1,\ell_2})}$ indicates concatenation of all outputs satisfying the path constraint $({\ell_1,\ell_2})\to(\ell_3 = 0)$. This step couples the scalar and equivariant latent spaces, where high-order structural information in the equivariant feature space is integrated into the scalar feature space of the nodes, further enhancing the structural representation capability of the node features.

Finally, Equation \ref{emb3} mixes the output results of all paths with the same output rotational order $\ell_3$ through an equivariant linear layer, generating representations of the same dimension as the input node equivariant features. This step compresses the output paths of the same rotational order into the same equivariant feature space. The parameters $w^{({\ell_1,\ell_2})\to(\ell), L}$ are trainable network parameters acting on different paths.

\textit{\textbf{Equivariant feature scalarization layer}}
To enhance the structural expressiveness of the model, this paper proposes a novel equivariant feature scalarization layer. This layer functions to transform equivariant features into scalar forms. This transformation not only extracts information from the  equivariant features of nodes but also provides structural constraints for the subsequent invariant feature transmission.

The input includes the existing equivariant features $\mathbf{\tilde{h}}_i^{(\ell),L}$ of the central particle $i$, the existing equivariant features $\mathbf{\tilde{h}}_j^{(\ell),L}$ of the neighboring particle $j$, and the equivariant message $\mathbf{\tilde{m}}_{ij}^{(\ell),L}$ of edge $(i,j)$. By using the C-G tensor product between these equivariant features, the layer outputs a set of scalar features for each edge $(i,j)$. The main computational steps are as follows:
\begin{align}
 &\mathbf{\alpha}_{ij}^{0,L} =  \mathop{\oplus}\limits_{({\ell_1,\ell_2})}\left(\mathbf{\tilde{h}}_i^{(\ell_1),L}  \mathop{\otimes}\limits_{{\ell_1,\ell_2}}^{\ell_3} \mathbf{\tilde{m}}_{ij}^{(\ell_2),L}\right)^{({\ell_1,\ell_2})\to(\ell_3 = 0),L} \label{inceq1}\\
 &\mathbf{\tilde{h}}_{ij}^{(\ell),L}  = \mathbf{\tilde{h}}_i^{(\ell),L} -\mathbf{\tilde{h}}_j^{(\ell),L} \label{inceq2}\\
 &\mathbf{\alpha}_{ij}^{1,L} =  \mathop{\oplus}\limits_{({\ell_1,\ell_2})}\left(\mathbf{\tilde{h}}_{ij}^{(\ell_1),L}  \mathop{\otimes}\limits_{{\ell_1,\ell_2}}^{\ell_3} \mathbf{\tilde{h}}_{ij}^{(\ell_2),L}  \right)^{({\ell_1,\ell_2})\to(\ell_3 = 0),L} \label{inceq3}.
\end{align}
As seen in Equations \ref{inceq1} and \ref{inceq3}, the scalarization of equivariant features mainly consists of two parts. Firstly, $\mathbf{\alpha}_{ij}^{0,L}$ is obtained through the interaction between $\mathbf{\tilde{h}}_i^{(\ell),L}$ and $\mathbf{\tilde{m}}_{ij}^{(\ell),L}$, reflecting high-order structural information of the current edge $(i,j)$ in the local structure. Secondly, for $\mathbf{\alpha}_{ij}^{1,L}$, the difference $\mathbf{\tilde{h}}_{ij}^{(\ell),L}$ between $\mathbf{\tilde{h}}_i^{(\ell),L}$ and $\mathbf{\tilde{h}}_j^{(\ell),L}$ is first calculated, indicating the structural correlation in the equivariant feature space between the central neighboring particles $i$ and $j$. The scalarization process in Equation \ref{inceq2} is completed through self-interaction of $\mathbf{\tilde{h}}_{ij}^{(\ell),L}$.

It is worth noting that Equations \ref{inceq1} and \ref{inceq2} retain only the scalar output of the tensor product. Therefore, during the tensor product calculation, the rotational orders $\ell_1$ and $\ell_2$ must satisfy $\ell_1 = \ell_2$. In this context, the self-interaction process in Equation \ref{inceq2} can also be written as $\mathbf{\alpha}_{ij}^{1,L} = \mathop{\oplus}_{({\ell})}\lVert \mathbf{\tilde{h}}_{ij}^{(\ell),L} \rVert$, where $\lVert \mathbf{\tilde{h}}_{ij}^{(\ell),L} \rVert$ represents the norm of the geometric tensor for rotational order $\ell$.

\textit{\textbf{Invariant message passing layer}}
This layer uses structural information extracted by the equivariant feature scalarization layer to further characterize the structure. A simple gated convolution mechanism is employed to obtain the final representation of the central node $i$:
\begin{align}
    &\mathbf{\alpha}_{ij}^{L}= \text{MLP}(\left[ \mathbf{\alpha}_{ij}^{0,L} \oplus \mathbf{\alpha}_{ij}^{1,L} \oplus \mathbf{e}_{ij}\right]) \label{ineq1}\\
    &\mathbf{h}_i^{L} = \Delta\mathbf{h}_i^L + \sum_{j\in\mathcal{N}_i}\mathbf{\alpha}_{ij}^{L} \odot \text{MLP}(\Delta\mathbf{h}_j^L)
\end{align}
In Equation \ref{ineq1}, all scalar features on the edge $(i,j)$ are mapped to structural constraint features on the edge, refining the correlation of nodes $i$ and $j$ in the local structure. These structural constraints are then used in the message passing process. After message aggregation is completed, the residual update in the scalar feature space transmits the scalar feature information from the equivariant message passing layer to the current layer. In experiments, it was found that compared to unconstrained invariant information propagation, the introduction of structural constraints effectively enhances the structural expressiveness.

\textit{\textbf{Graph normalization}}
After completing the equivariant and invariant message passing processes, the graph normalization layer normalizes the nodes' invariant and equivariant features to stabilize training. For invariant features, graph normalization is used \cite{cai2021graphnorm}. To ensure feature equivariance, the normalization of equivariant features needs to be conducted independently for different orders of features \cite{geiger2022e3nn}. Specifically, the normalization process for invariant features is as follows:
\begin{equation}
    \mathbf{{h}}_i^L  = \frac{\mathbf{{h}}_i^L - \eta\odot{{\rm{E}}[\mathbf{{h}}_i^L]}} {\sqrt{{\rm{Var}^L}[\mathbf{{h}}_i^L - \mathbf{\eta}\odot{{\rm{E}}\left[\mathbf{{h}}_i^L\right]}]}} \odot \mathbf{\beta^L} + \gamma^L 
\end{equation}

where E and Var represent batch mean and variance, respectively. $\eta^L, \beta^L,$ and $\gamma^L$ are learnable parameters. The normalization of equivariant features is based on the norm of each rotational order:
\begin{equation}
    \mathbf{\tilde{h}}_i^{(\ell.L} = \frac{\mathbf{\tilde{{h}}}_i^{(\ell),L}}{{\sqrt{{\rm{E}}\left[||\mathbf{\tilde{h}}_i^{(\ell),L}||^2\right]}/\sqrt{2\ell+1}}} \odot\gamma^{(\ell)},
\end{equation}
where $||\mathbf{{h}}_i^{(\ell),L}||$ is the norm of the rotational order $\ell$ feature, and $\gamma^{(\ell), L}$ is a learnable weight. This normalization method ensures that features maintain their equivariance during training by independently processing equivariant features of different rotational orders.

\textit{\textbf{Graph decoder}}
After passing through $N_L$ independent structure representation layers, the final invariant structural representation of the nodes is obtained and fed into the graph decoder. The EIGNN graph decoder consists of a subgraph readout layer and an MLP layer for regressing the target, yielding the final dynamic prediction results. The subgraph readout layer perform an additional meassage passing on the subgraph to further refine the invariant feature representation:

\begin{equation}
     \mathbf{{h}}_i^{out} = \sum_{j \ in \mathcal{N_i}}\text{MLP} \left(\mathbf{{h}}_i^{L} \oplus \left( \mathbf{{h}}_i^{L} -\mathbf{{h}}_j^{L} \right)\right) 
\end{equation}
The subsequent MLP layer then operates on $\mathbf{{h}}_i^{out}$ to generate the predictions.

\begin{table*}[!ht]  
    \centering  
    \caption{Comparison of Computational Complexity of Tensor Products in SE(3)-GNN \cite{pezzicoli2024rotation} and EIGNN}  
    \label{tab:seduibiei}  
    \begin{tabular}{cc|cc}  
        \toprule[1.5pt]
        & SE(3)-GNN \cite{pezzicoli2024rotation} & \multicolumn{2}{c}{EIGNN} \\  
        \midrule[1.0pt]
        Interaction Mode & Edge-wise & Node-wise & Edge-wise \\  
        Tensor Product ($a \otimes b$) & Fully-connected  & Fully-connected  & Element-wise  \\  
        Channel number (C) & 6 & 1 & 1 \\
        Number of Tensor Products ($N_t$) & $|\mathcal{E}|$ & $N_p$ & $|\mathcal{E}|$\\
        Complexity ($O(N_t \cdot C \cdot \ell_{max}^6)$) & $O(|\mathcal{E}| \cdot 6 \cdot \ell_{max}^6)$ & $O(N_p \cdot\ell_{max}^6)$ & $O(|\mathcal{E}| \cdot \ell_{max}^6)$\\  
        \bottomrule[1.5pt]  
    \end{tabular}  
\end{table*}

\subsection{ Complexity Analysis}

Equivariant GNNs based on spherical harmonics require C-G tensor products for feature interactions. The high computational cost of tensor products hinders their application to large-scale glassy data. The proposed equivariant GNN, which incorporates invariant message passing under equivariant message constraints, reduces the dependency on C-G tensor products compared to the standard equivariant message passing networks like SE(3)-GNN\cite{pezzicoli2024rotation}. 

Table \ref{tab:seduibiei} compares the computational complexity of tensor products in EIGNN and SE(3)-GNN. The tensor products in EIGNN primarily arise from the equivariant message passing layer and the equivariant feature scalarization layer. In SE(3)-GNN, the fully-connected tensor products are performed on the edges, whereas in EIGNN, they are executed at the nodes, which significantly reduces the number of tensor products since the number of nodes is far less than the number of edges. Although the scalarization layer performs tensor products on the edges, the use of single-channel equivariant features, along with the computation of only scalar outputs from the tensor products, makes this process lightweight. Overall, the proposed EIGNN model has lower computational complexity than SE(3)-GNN.

\section{Results}
\subsection{Experiment setup}
In this section, we further analyze the performance of EIGNN in predicting the dynamic properties of glassy system. 

\textit{{\textbf{Baseline}}}
This study compares our approach with all models in the GlassBench-3D benchmark. These include GlassMLP \cite{jung2023predicting}, CAGE \cite{alkemade2023improving}, BOTAN \cite{shiba2023botan}, and SE(3)-GNN \cite{pezzicoli2024rotation}. Experimental results for the benchmarked models are sourced from the GlassBench-3D benchmark \cite{jung2023roadmap}. 

\textit{{\textbf{Paramters}}}
The study employs a network architecture consisting of 7 interaction layers. The dimensions for both the node invariant features and intermediate invariant features in the interaction layers are uniformly set to 32. The maximum rotational order of the equivariant features, $\ell_{max}$, is set to 2.

\textit{{\textbf{Loss Function}}}
This study employs the Mean Square Error (MSE) as the optimization criterion, measuring the discrepancy between the actual and predicted propensity of type A paritcles. In line with the SE(3)-GNN model, this work adopts multivariate regression, simultaneously regressing the dynamic propensity  at all timescales. This approach avoids the additional overhead of training a separate model for each timescale.

We also find the target normalization is important for network performance in multivariate regression settings. Due to the substantial variability in the dynamical propensity of particles across different time scales, additional bias during training will introduced without preprocessing.  Therefore, the loss function is defined as follows:
\begin{equation}
\text{Loss} = \sum_{t=1}^{N_t} \sum_{i=1}^{N_{\text{batch}}} \left\lVert {\hat{y}}_i(t) - \left(\frac{{y}_i(t) - \mu_{\rm batch}(t)}{\sigma_{\rm batch}(t)}\right) \right\rVert^2,
\end{equation}
where, $N_t$ denotes the number of time points sampled during the relaxation process, and $N_{\text{batch}}$ represents the total number of particles in the current training batch. ${\hat{y}_i(t)}$ is the predicted propensity for particle $i$ at time $t$, while ${y}_i(t)$ is the actual value. We normalize the targets at each time point by the mean $\mu_{\rm batch}(t)$ and the standard deviation $\sigma_{\rm batch}(t)$ of the actual mobility of all particles within each training batch.

\textit{{\textbf{Evaluation Criteria}}}
Here, we use two different calculation strategy of Pearson correlation coefficients. The first one  employed the mean of Pearson correlation coefficients across all test configurations, denoted as Pcc(mean). Another one directly calculates the Pearson correlation between the predicted and actual values for all particles in all configurations, denoted as Pcc(all). The formula for Pcc(all) is as follows:
\begin{align}
&    \text{Pcc(mean)} =\frac{1}{N_{\text{test}}} \sum_{i =1}^{N_{\text{test}}}\frac{\text{Cov}(\hat{\mathbf{y}}_i(t), \mathbf{y}_i(t))}{\sqrt{\text{Var}(\hat{\mathbf{y}}_i(t)) \cdot\text{Var}(\mathbf{y}_i(t)})} \\
&\text{Pcc(all)} = \frac{\text{Cov}\left(\hat{\mathcal{Y}}(t), \mathcal{Y}(t)\right)}{\sqrt{\text{Var}\left(\hat{\mathcal{Y}}(t)\right) \cdot \text{Var}\left(\mathcal{Y}(t)\right)}}
\end{align}
In this equation, $N_{\text{test}}$ denotes the number of test graphs, Cov represents the covariance between variables, and Var denotes the variance of a variable. $\hat{\mathbf{y}}_i(t)$ and  ${\mathbf{y}}_i(t)$ is the predicted and actual propensity of $i$-th configurations, respectivley. $\hat{\mathcal{Y}}(t) =  \left[ \hat{\mathbf{y}}_1(t) \oplus ... \oplus \hat{\mathbf{y}}_{N_{\text{test}}}(t)\right]$ and $\mathcal{Y} (t)= \left[ \mathbf{y}_1(t) \oplus ... \oplus\mathbf{y}_{N_{\text{test}}}(t)\right]$ are the concatenated vectors of predicted and actual values for all configurations in the test set, respectively.

\textit{\textbf{Training and test}}
The proposed EIGNN model is built and trained by the PyTorch Geometric \cite{fey2019fast} and e3nn library\cite{geiger2022e3nn}. 
Depending on the equilibrium positions used by EIGNN, we denote the model that utilizes inherent structure as input as EIGNN-Inh, and the one using cage state positions as EIGNN-Cage. 
Each model is trained and tested separately. For EIGNN-Inh, the dataset comprises 400 configurations for training and 100 configurations for testing. For EIGNN-Cage, 100 configurations are used for both training and testing. Each configuration provides 10 relaxation dynamics labels, and this study compares data from the 2nd to the 9th time periods, covering the initial cage motion at $t = 1.3$ to the diffusion motion at $t = 41200$.

During training, we employed the Adam optimizer with an initial learning rate set to 0.001. We fixed the training epoch to 100 with one graphs in the training batch. Additionally, we utilized the StepLR algorithm to adjust the learning rate at fixed epoch intervals. The training and testing of the EIGNN models in this study were conducted using an A100 GPU.

It is worth noting that due to the use of graph regularization and target regularization in EIGNN, using a single test graph may introduce some bias when measuring the Pcc(all) metric. To mitigate this bias, we employ batch processing techniques, combining 10 test graphs into one batch for testing when measuring Pcc(all).
\begin{figure*}[ht]
    \centering
    \includegraphics{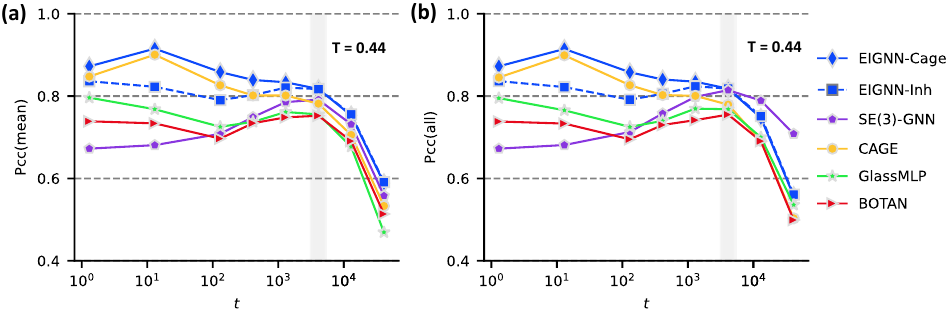}
    \caption{\textbf{(a)} Pcc(mean) and \textbf{(b} Pcc(all) results of various models at temperature T=0.44. The shaded area indicates the region near the relaxation time.}
    \label{fig:t044}
\end{figure*}
\begin{figure*}[!ht]
    \centering
    \includegraphics{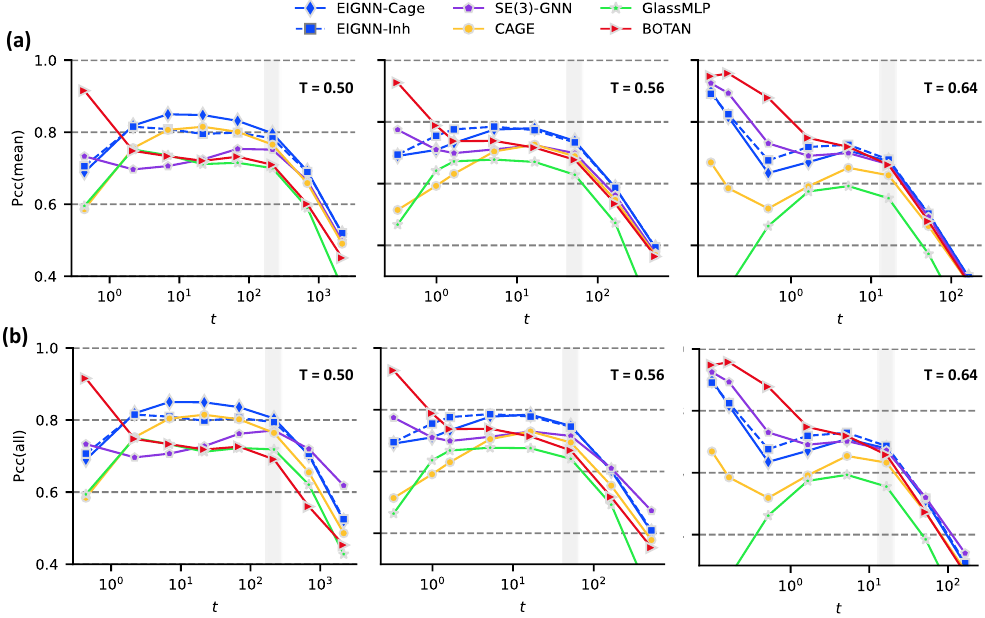}
    \caption{\textbf{(a)} Pcc(mean) and \textbf{(b} Pcc(all) results of various models at temperature T=0.50, 0.56 and 0.64.}
    \label{fig:t064}
\end{figure*}
\subsection{Comparative analysis of different models}
\textit{\textbf{Results on temperature T=0.44 }}
A comparative analysis of various models at temperature T=0.44 is depicted in Figure \ref{fig:t044}. It is evident from the figure that the EIGNN model utilizing the cage-state structure achieves optimal results across two metrics from the initial stages of cage motion up to the relaxation time scale. The CAGE model, which also employs the cage-state structure, achieve the second best in the short time. In contrast, other models using the inherent structure significantly underperform in short-time predictions compared to the CAGE and EIGNN-Cage models. This further substantiates the significance of incorporating the cage-state structure in short-time predictions. This model can effectively encode the directional feature of particles moving towards the cage center, whereas the CAGE model can only handle scalar features, hence the proposed model demonstrates more precise predictions. 

The results presented in \cite{pezzicoli2024rotation} claim that the information between the initial state and inherent state is not beneficial to the performance of the SE(3)-GNN model. However, the results of the EIGNN-Inh model demonstrate that the displacement between the initial state and inherent state plays an important role in propensity prediction.

In the long time, the EIGNN-Cage and EIGNN-Inh models exhibit comparable predictive outcomes, indicating that both structures play a similar role in long-time particle predictions. As the influence of the original cage-state structure on long-time dynamics diminishes after a longer duration beyond the relaxation time, the performance of all models declines.  In the Pcc(all) metric, the SE(3)-GNN model surpasses the models proposed in this paper in predicting dynamics after the relaxation time. However, the proposed EIGNN-Cage and EIGNN-Inh models still outperform all other benchmark models from the relaxation time in the Pcc(mean) metric.

\textit{\textbf{Results on higher temperatures }}
In this section, we delve into the performance of various models under high-temperature conditions. Figures \ref{fig:t064} illustrate the Pcc(mean) and Pcc(all) for the predictive outcomes at temperatures T=0.50, 0.56, and 0.64. The EIGNN-Cage and EIGNN-Inh models have demonstrated satisfactory predictive performance, particularly at temperatures T=0.50 and 0.56, where they outperform other existing methods across most time scales.

Unlike the results at T=0.44, we observe significant differences in the dependency of various models on inherent and cage structures at other temperature conditions. As the temperature gradually increases, the models' reliance on these two structures for short-time relaxation dynamics predictions diminishes. As depicted in the figures, on short time scales, the BOTAN model, which utilizes only the initial thermal noise structure, shows increasingly enhanced predictive performance. Particularly at the highest temperature T=0.64, models that depend on inherent and cage structures do not perform as well as the BOTAN and SE(3)-GNN models in short-time dynamics predictions prior to the relaxation time.

This phenomenon may be attributed to significant changes in particle motion patterns at high temperatures. As the temperature rises, the residence time associated with cage motion gradually decreases. At T=0.64, the primary modes of particle motion shift to short-duration hopping and long-duration diffusion, with cage motion becoming less pronounced. Consequently, the dependency on cage structure also decreases. Therefore, the performance of the EIGNN-Inh model, which utilizes the inherent structure, gradually surpasses that of the EIGNN-Cage model as the temperature increases.

\begin{figure*}
    \centering
    \includegraphics{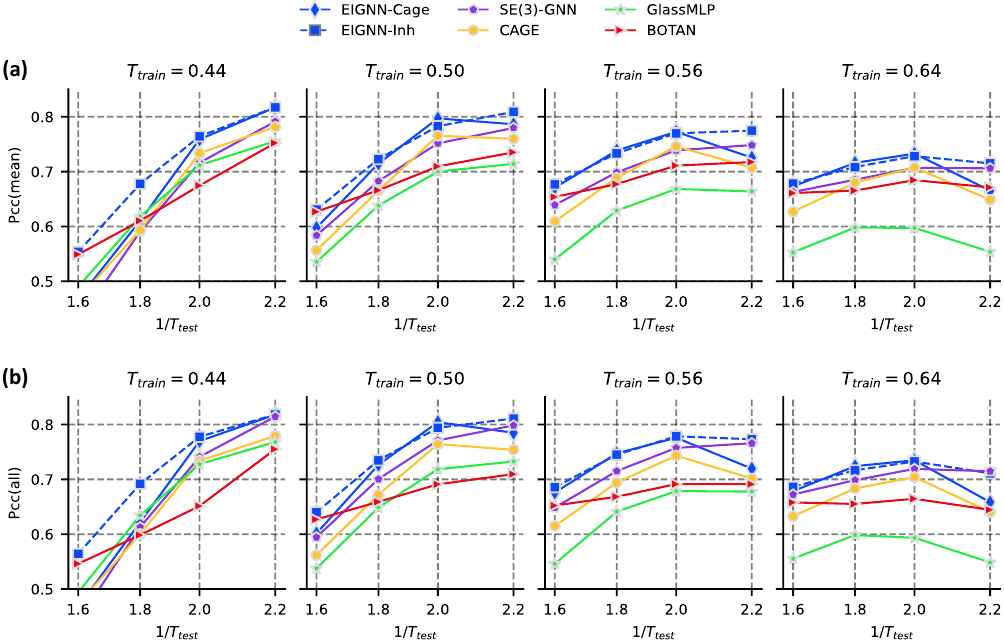}
    \caption{Pcc(mean) and Pcc(all) results of the model trained at temperature $T_{train}$ transferred to other temperatures.}
    \label{fig:transfer}
\end{figure*}
\subsection{ Transferability to different temperatures}

\textit{\textbf{Training with  separate temperature }}
FIrstly, we investigate the models' transferability, aiming to ascertain whether a model trained at a specific temperature can predict relaxation dynamics at other temperatures. 

Figure \ref{fig:t044} illustrates the models' ability to transfer to other temperatures using the Pcc(mean) and Pcc(all) metrics, respectively. The results reveal that the proposed models  exhibit the most outstanding temperature transferability, showing a significant advantage over existing models.

The EIGNN-Inh model, which utilizes the inherent structure, demonstrates superior temperature transferability compared to the EIGNN-Cage model that employs the cage-state structure. The primary reason for the lower transferability of the EIGNN-Cage model compared to the EIGNN-Inh model is the temperature-dependent properties of the cage-state structure. The equilibrium positions obtained at a specific temperature have a strong temperature bias, leading to a decline in performance when the EIGNN-Cage model is transferred to other temperatures. This phenomena is in contrast to the results exhibited by the CAGE model, which uses the cage-state structure, and the GlassMLP model, which uses the inherent structure. This may due to the suboptimal structural representation capability of the GlassMLP. Moreover, the trained EIGNN-cage models and CAGE models declines in their performance when transfer to the lower temperature T = 0.44 , further demonstrated the temperature-dependent properties of the cage-state structure. Furthermore

Although the SE(3)-GNN model, trained at T=0.44, achieves predictive performance comparable to the EIGNN models at that temperature, it exhibits poor transferability from low to high temperatures. This suggests that the SE(3)-GNN model is more susceptible to overfitting at low temperatures. In contrast, the proposed EIGNN models, especially the EIGNN-Inh, demonstrate enhanced transferability by avoiding overfitting to dynamic behaviors at specific temperatures. Furthermore, the EIGNN-Inh model, trained at a higher temperature of T = 0.50, exhibits performance nearly equivalent to that of an EIGNN model trained directly at the lower temperature of T = 0.44 when predicting the relaxation dynamics at T = 0.44. 

overall, the EIGNN-Inh model, trained at even higher temperatures, can effectively transfer to predict relaxation dynamics at lower temperatures. These results further confirm that the structure-dynamics relationship is relatively stable within a certain temperature range, providing an effective approach to explore and predict their behavior under extreme conditions through machine learning models.
\begin{figure}[t]
    \centering
    \includegraphics{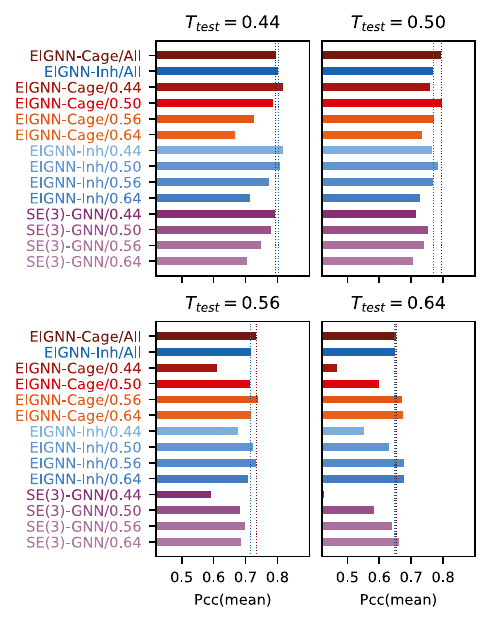}
    \caption{Pcc(mean) and Pcc(all) results of the models trained with the combined dataset from all temperatures.}
    \label{fig:temp_all}
\end{figure}

\textit{\textbf{Training with all temperature}} An intriguing question arises: Can we enhance the  transferability of the model by training on data from all temperatures simultaneously? To further investigate this question, we retrain the EIGNN model with the data covering all temperatures. Since there are 100 training configurations with cage-state structures at each temperature, the combined training set for EIGNN-cage would consist of 400 configurations. For the combined training set of EIGNN-Inh, we randomly selects 100 configurations from  training set of each temperature to reduce the training data size. Consequently, the number of configurations for training both the EIGNN-Cage and EIGNN-Inh models is 400 configurations.


Figure \ref{fig:temp_all} presents the performance of various models when trained on the combined temperature data. In the figure, the text before and after "/" represents the model name and training temperature, respectively. For example, "EIGNN-Inh/All" represents the model trained on the dataset combining all temperatures, while "EIGNN-Inh/0.44" represents the model trained on the complete dataset at T=0.44.

It can be observed that the performance of EIGNN-Cage/All and EIGNN-Inh/All in individual tests at different temperatures is slightly lower than that of the EIGNN models trained at the specific temperature. This is reasonable because the combined dataset includes dynamics at various temperatures, making the correlation between structure and dynamics more complex and thus leading to a slight decrease in model performance. However, it is noteworthy that EIGNN-Cage and EIGNN-Inh still outperform most of the transfer test models. For example, in the test results at the lowest temperature T=0.44, the prediction results of EIGNN-Cage/All are significantly higher than those of the models trained at T=0.50, 0.56, and 0.64. Similarly, the prediction results of the mixed-trained EIGNN-Inh/All are also higher than those of the models trained at T=0.56 and 0.64. Furthermore, EIGNN-Cage/All and EIGNN-Inh/All outperform the SE(3)-GNN models trained at those temperatures and the transfer models at test temperatures T=0.44, 0.50, and 0.56.

In summary, training the model on a new dataset formed by combining different temperatures still achieves satisfactory dynamics prediction results. These findings strongly suggest that our model not only learns complex structural features related to dynamics but also avoids overfitting dynamic behaviors at specific temperatures. This further indicates that the dynamic heterogeneity at different temperatures in glassy materials originates from the same structural characteristics.
\begin{figure}[!t]
    \centering
    \includegraphics{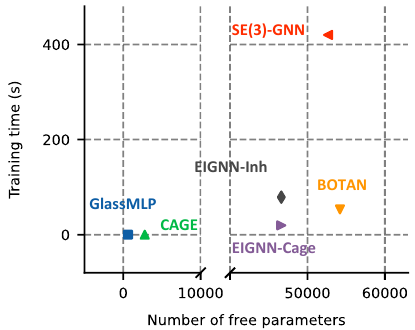}
    \caption{comparative analysis of the number of parameters and training times across different models. Due to the short training time of the Cage and GlassMLP models, their training times are approximately zero.}
    \label{fig:time}
\end{figure}
\begin{figure*}[ht]
\centering
\includegraphics{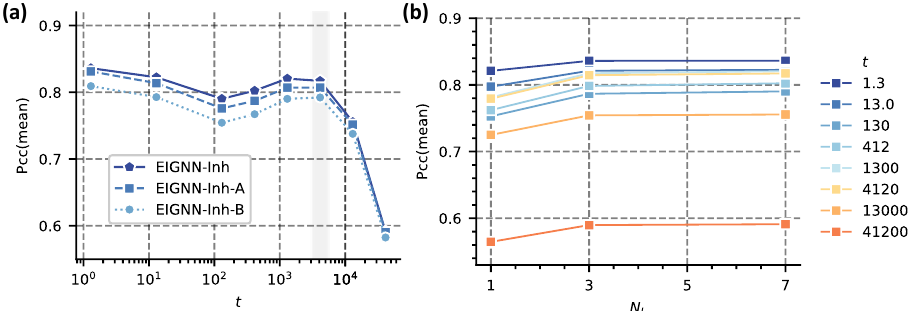}
\caption{\textbf{(a)} Prediction results of EIGNN-Inh-A and EIGNN-Inh-B at temperature T=0.44.\textbf{(b)}Prediction results of EIGNN-Inh models with different number of interaction layer at temperature T = 0.44}
\label{fig:aba}
\end{figure*}

\section{Further analysis}
\subsection{Scalability analysis}
Figure \ref{fig:time} provides a detailed comparison of the number of training parameters and training times of various models. GlassMLP and CAGE utilize traditional machine learning models and have the fewest parameters and training time. However, the preporcessing of the physical descirptors are still complexity and time-costing. In constraint, the GNN models input with only the coordinates are more flexible. Comparison with SE(3)-GNN and BOTAN, EIGNN has the fewest parameters. BOTAN has the highest parameter count among the models. 

 In terms of training time, EIGNN demonstrates significant advantages over traditional equivariant graph neural network. The dependency of SE(3)-GNN on equivariant feature interactions results in a substantial increase in training time. In contrast, the proposed equivariant-constrained invariant message passing mechanism  significantly reduces computational complexity and thereby decreases training time. Additionally, EIGNN-Cage exhibits the shortest training time, primarily because this model uses only a quarter of the configurations based on cage structures compared to inherent configurations. Moreover, EIGNN-Inh shows a modest increase in training time compared to the non-equivariant-constrained BoTAN model. This further underscores the efficiency of the proposed model in establishing the relationship between structure and dynamics.
\subsection{Ablation analysis}
\textit{\textbf{The impact of equiviarnt feature scalarization}} The introduction of equivariant information in EIGNN is achieved through parameters $\alpha_0$ and $\alpha_1$ in the equivariant information scalarization layer. The structural parameter $\alpha_0$ can project the central node features onto the edge features, while $\alpha_1$ scalarizes the equivariant features of the central node and its neighboring nodes. These two parameters are the key to encode the displacement vectors and geometric features.

To further validate the impact of these parameters, two variants of EIGNN-Inh, named EIGNN-Inh-A and EIGNN-Inh-B were designed. EIGNN-Inh-A removes the structural parameter $\alpha_0$, retaining only $\alpha_1$, while EIGNN-Inh-B  removes both parameters, resulting in a model that contains only invariant features and the invariant feature message-passing layer.

Figure \ref{fig:aba}(a) shows the prediction results of EIGNN-Inh-A and EIGNN-Inh-B at temperature T=0.44.  It can be observed the lack of structural parameter $\alpha_0$ in EIGNN-Inh-A affects its performance from the early stage of diffusion dynamics to the relaxation time. This indicates that the $\alpha_0$ plays a crucial role in the prediction of relaxation dynamics. Further observation of the prediction results of EIGNN-Inh-B reveals that the absence of encoding equivariant features also affects the prediction performance of the model across multiple time scales. Here, EIGNN-Inh-B includes only the magnitude of the displacement vectors and scalar geometric features, and overlook the directional component. This may hinder a complete depiction of the intricate particle interactions characteristic of the cage effect dynamics, which may consequently affect the efficacy of models in correlating structural features with dynamic behaviors.

\textit{\textbf{The impact of the number of interaction layer}} Figure \ref{fig:aba}(b) illustrates the prediction results of EIGNN-Inh models with different number of interaction layer at temperature T = 0.44. Even with a single interaction layer, the model maintains a high prediction accuracy. When three interaction layers are employed, the performance significantly improves compared to the single-layer model. Additionally, it is observed that the EIGNN-Inh model with three interaction layers achieves a limited increase in accuracy compared to the model with seven layers.

These results further demonstrate the structural encoding capability of the model. The three-layer message-passing mechanism has already captured essential structural features, providing a robust tool for exploring the dynamics of glassy structures with a lightweight EIGNN-Inh model.
\begin{figure*}[htp]
    \centering
    \includegraphics{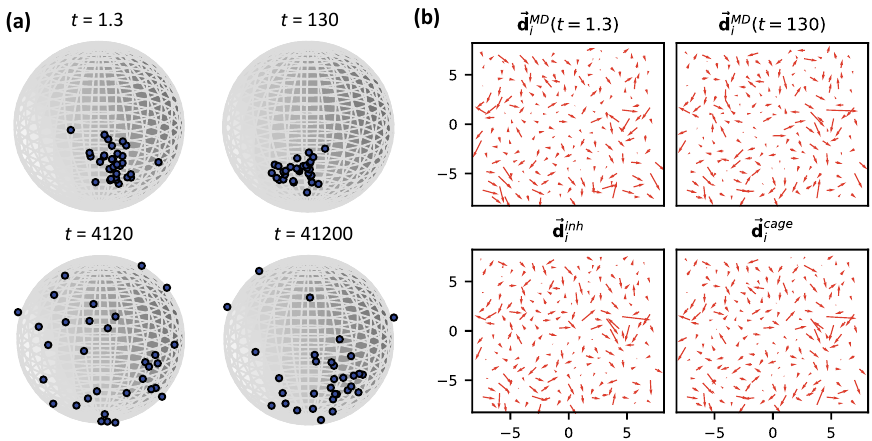}
    \caption{\textbf{(a)} The distribution of the movement directions of a single particle across multiple independent relaxation simulations at t= 1.3, 130, 4120 and 41200.\textbf{(b)} Spatial distribution of the tendency of particle displacements at t = 1.3 and t=130 and the displacement parameters based on the inherent state and cage state on a 2d slice ($3.9<z<5.1$).}
    \label{fig:phya}
\end{figure*}
\subsection{The analysis of the Displacement parameter}

It is worth further investigating why this parameter is effective. Previous studies have focused only on the propensity of particles, i.e., the average displacement of particles, while neglecting the direction of relaxation motion. Recent literature suggests that there is a certain directionality in the way particles vibrate in their cagec \cite{PhysRevLett.122.015501} or escape from their cages \cite{annamareddy2022distribution}. To observe the directional tendency of particle motion, Figure \ref{fig:phya}(a) visualizes the distribution of movement directions of one particle at different times. It can be seen that the particle will move in various directions during independent relaxation processes. However, at the cage motion times of t=1.3 and t=130, the movement directions of the particles are relatively concentrated, primarily towards a fixed direction. This indicates a tendency for particles to move in a particular direction during cage motion. However, at the relaxation time t=4120, the movement directions of the particles become more dispersed. 

Previous studies have shown that the distance between the equilibrium structure and the initial structure of particles can effectively capture the propensity of particles. But can the displacement parameter capture the directional tendency of particle cage relaxation dynamics? To address this question, we can define the tendency of particle displacement as $\mathbf{\vec{d}}_i^{\rm MD}(t) = <\mathbf{\vec{r}}_i(t) - \mathbf{\vec{r}}_i^{\rm init} >_{\rm  iso}$.

Figure \ref{fig:aba}(b) compare the spatial distribution with the displacement parameters based on the inherent state and cage state with the tendency of particle displacement at times t=1.3 and t=130. Interestingly, the correspondence among these four parameters is quite good, suggesting that the displacement parameter defined based on inherent and cage structures effectively captures the directional tendency in short-time dynamics. Consequently, the directional information enhanced the prediction capability of our proposed EIGNN model.
\section{Conclusion}
This paper introduces a novel directional order parameter that is defined based on the equilibrium positions of particles within a system. This parameter provides a comprehensive description of the magnitude and direction of the deviation of the initial configuration of the particles from their equilibrium positions. To effectively leverage this parameter, an equivariant graph neural network (EIGNN) is designed to simultaneously represent both the physical parameter and the geometric structure of the particles. The EIGNN is able to capture the intricate relationships between the directional order parameter and the geometric structure, enabling accurate predictions of the dynamic behavior of the particles. Experimental results demonstrate that the proposed model exhibits excellent dynamic prediction capabilities and temperature transferability across various time periods and temperatures. These findings further validate the effectiveness of the directional order parameter in establishing a strong correlation between the structure and dynamics of the particles, and highlight the potential of the EIGNN for modeling and predicting complex physical systems.


\bibliographystyle{unsrtnat}

\bibliography{ref}

\end{document}